\documentclass[twocolumn,aps,prl,showpacs
]{revtex4}
\usepackage{graphicx}
\usepackage{epsf}
\usepackage{amsmath}
\usepackage{amssymb}

\DeclareMathAlphabet{\mathpzc}{OT1}{pzc}{m}{it}

\begin{document}

\title{A topological classification of interaction-driven spin pumps}
\author{Dganit Meidan}
\affiliation{Dahlem Center for Complex Quantum Systems and Institut
f\"{u}r Theoretische Physik, Freie Universit\"{a}t Berlin, 14195
Berlin, Germany}

\author{Tobias Micklitz}
\affiliation{Dahlem Center for Complex Quantum Systems and Institut
f\"{u}r Theoretische Physik, Freie Universit\"{a}t Berlin, 14195
Berlin, Germany}

\author{Piet W.~Brouwer}
\affiliation{Dahlem Center for Complex Quantum Systems and Institut
f\"{u}r Theoretische Physik, Freie Universit\"{a}t Berlin, 14195
Berlin, Germany}

\date{\today} 

\begin{abstract}
When adiabatically varied in time,  
certain one-dimensional band insulators allow for the quantized noiseless pumping 
of spin even in the presence of strong spin orbit scattering.
These spin pumps are closely related to the 
quantum spin Hall system, and their properties are protected by
a time-reversal restriction on the pumping cycle.
In this paper we study pumps formed of one-dimensional insulators  
with a time-reversal restriction on the pumping cycle and 
a bulk energy gap which arises due to interactions. 
We find that the correlated gapped phase can lead to novel pumping properties. In particular, systems with $d$  different 
ground states 
can give rise to $d+1$ different classes of spin pumps,  
including a trivial class which does not pump quantized spin and
$d$ non-trivial classes allowing for the pumping of quantized spin $\hbar/n $ on average per cycle, where $1\leq n\leq d$. 
We discuss an example of a spin pump that transfers on average spin $ \hbar/2$ 
 without transferring charge. 
\end{abstract}

\maketitle

{\it Introduction:---}
Finding ways to manipulate individual charges or spins at zero external bias, the idea that lies at the essence of pumping, holds promise for numerous applications.
In a seminal work, Thouless observed that certain band insulators allow for the  adiabatic 
pumping of quantized charge~\cite{DJThoulessPRB1983}. Transcending  
its relevance for practical applications, this observation sheds light on  transport properties found in other systems. 
Notably, Laughlin's argument for the quantization 
of Hall conductance~\cite{RBLaughlinPRB1981} can be formulated in terms of a quantized charge pump~\cite{QNiuJPA1984}. In accordance with the quantum Hall system,  
the adiabatic charge pump formed of a band insulator 
can be characterized by a $Z$ topological 
invariant, which determines the quantized charge pumped in one cycle. 

These ideas have found a recent  extension to spin pumps with
the discovery of the quantum spin Hall effect in
time-reversal invariant systems~\cite{LFuPRB2006}. 
In analogy to the quantum Hall state, it is possible to 
gain insight into the spin Hall state 
by studying a pump formed by placing  the two-dimensional  system on a cylinder with a circumference of a single unit cell and threading it with a magnetic flux which is varied in time~\cite{DJThoulessPRB1983,QNiuJPA1984}. 
Time-reversal symmetry of the two-dimensional Hamiltonian imposes a time-reversal restriction on the pumping cycle~\cite{LFuPRB2006}. 
Similarly to its two-dimensional analog, spin pumps with a time-reversal restriction on their pumping cycle are characterized by a $Z_2 $ 
topological invariant.  
The non-trivial class of pumps allows for the symmetry-protected pumping 
of quantized spin, even in the presence of strong spin orbit scattering~\cite{DMeidanPRB2010}. 
Following the paradigm of the fractional quantum Hall effect~\cite{RTaoPRB1984,QNiuPRB1985,DJThoulessPRB1989,XQWenPRB1990}, can interactions give rise 
to spin pumping properties that can not be found in non-interacting pumps?

In this paper we study the topological classification of spin pumps consisting of a family of one-dimensional insulators  in which the bulk gap arises due to electron-electron interactions. Our classification is made with respect to the observable pumping properties of pumps that are weakly coupled to leads, not on the structure of the bulk insulating state. We find that the number of classes in the correlated system is larger than in the non-interacting case if the system has $ d$  different many body ground states. In particular, a spin pump with $d$ ground states gives rise to $d+1$ distinct classes which exhibit different pumping properties: For a weakly coupled pump, these are a trivial class, which does not pump quantized spin, and $d $ non-trivial classes. The non-trivial classes include an integer spin pump that allows for the pumping of quantized spin $\hbar$  during a cycle, as well as $d-1$ ``fractional'' spin pumps that allow for the average pumping of quantized spin $\hbar/n$ with $1< n \leq d$ per cycle. We discuss an example of a pump that transfers an average of spin $ \hbar/2$ during a pumping cycle, without transferring charge.

{\it $Z_2$ classification of non-interacting spin pumps:---}
To set the stage, we briefly review 
the $Z_2$ classification of non-interacting spin pumps. We then show that this classification is naturally extended to interacting systems.
We consider a family of one-dimensional 
Hamiltonians with a bulk energy gap 
that depend continuously on a cyclic pumping parameter 
$t$  and satisfy
\begin{align}
\label{time-reversal transformations}
H(t+T)=H(t),\qquad  H(-t)=\Theta H(t)\Theta^{-1},
\end{align} where $ \Theta$ is the time-reversal operator. We assume that $H(t)$ does not posses any additional discrete symmetries. Due to the time-reversal restriction \eqref{time-reversal transformations}, such system cannot pump charge but it may pump spin. These pumps are related to quantum spin Hall systems (class AII in the classification of Ref.~\onlinecite{AAltlandPRB1997}) upon placing the two dimensional system on a cylinder threaded by magnetic flux. [This connection becomes evident upon the identification $(k_x,k_y)\to(k_x,t)$ \cite{LFuPRB2006}.] 

When coupling the system to one-dimensional non-interacting leads, the transport properties of the open system at time $t $ are  
determined from the scattering matrix. Provided that the system's size exceeds the attenuation length associated with 
the bulk energy gap, the scattering matrix decouples into two unitary $2\times 2$ 
reflection matrices ${\cal R}_\alpha$, for left and right leads $\alpha=L/R$, respectively. 
The average spin injected into lead $\alpha$ during the  pumping cycle~\cite{PWBrouwerPRB1998}, 
$\vec{S}_\alpha 
=  {\hbar \over 2\pi} \int_0^T dt\,
{\rm Im}\, \textrm{tr}
\left([d{\cal R}_\alpha/dt] \vec{\sigma}{\cal R}_\alpha^\dag\right)$ is invariant under a $ U(1)$ gauge transformation and 
depends only on the particle-hole symmetric $SU(2)=U(2)/U(1)\simeq S^3$ part of 
the reflection matrix, denoted by $\tilde{\cal R}$, 
where we have dropped the lead index $\alpha$ for brevity.
Hence, the pumping cycle can be visualized as  a loop which 
$\tilde{\cal R}(t)$ forms on the three sphere $S^3$.

The symmetry constraints~\eqref{time-reversal transformations} 
lead to  similar constraints on the reflection matrix
\begin{align}\label{niconstr}
\tilde{\cal R}(t) 
&= \sigma_2\tilde{\cal R}^T(-t)\sigma_2,
\qquad
\tilde{\cal R}(t+T)
= \tilde{\cal R}(t).
\end{align} 
These constrains ensure the existence of two time-reversal invariant moments (TRIM) $t_1=0$ and $t_2=T/2$, at which $\tilde{\cal R}(t_i)=\pm\openone$, which corresponds to the occurrence or absence of a pair of resonances, that occur precisely at the TRIM in the presence of particle-hole symmetry, which fixes the $U(1)$ part of ${\cal R}$ to be unity. (For a generic $ U(1)$ phase, the pair is symmetrically split around the TRIM and related by time-reversal)~\cite{CommentAccResonance}.
Following Ref.~\cite{DMeidanPRB2010}, the parity of resonance pairs around the TRIMs defines a $Z_2$ index:
\begin{align}
\label{z2_index}
\tilde{\cal R}(t_1)\tilde{\cal R}(t_2)=\pm\openone^{z_2}.
\end{align}
Any loop $\tilde{R}_\alpha(t)$ on the three sphere characterized by $z_2=0 $ can be contracted onto a single point and, hence, corresponds to a trivial pump. Alternatively, paths with $z_2=1$ cannot be contracted \cite{DMeidanPRB2010}.  

{\it Generalization to interacting systems:---}
Extending the considerations presented above to systems in which the gap arises due to many body interactions requires that the description of transport in terms of a unitary scattering matrix remains meaningful. As interactions may lead to inelastic scattering, the unitarity of the scattering matrix is not ascertained in general. 
In the presence of a finite energy gap $\Delta$ for bulk excitations, bulk charge and spin excitations are absent at sufficiently low temperatures $\beta^{-1}\ll\Delta $. Nonetheless, inelastic scattering can still arise due to ground state degeneracies or mid-gap states at the edge of the wire. 
Inelastic scattering involving the excitation of boundary states at energy $\epsilon < \Delta$ are suppressed in the weak coupling limit $\Gamma\ll\epsilon$.
A more subtle effect may arise in the presence of (Kramers) degenerate edge states, where a Kondo effect may develop. For the purpose of classification, we may restrict to  degeneracies protected by time-reversal symmetry. Such degenerate states generically occur at a finite distance $\mu$ from the Fermi energy in the leads if there are no additional discrete symmetries (such as particle-hole symmetry). This leads to an exponentially small Kondo temperature $\beta_{\rm K}^{-1} \sim e^{-\mu/\Gamma}$ \cite{ACHewson1993}. In the perturbative limit, $\beta_{\rm K} \gg \beta$, transitions between the Kramers degenerate pair occur at a rate $\Gamma^2/(\mu^2\nu_0) \ll \Gamma$, where $\nu_0$ is the density of states in the lead. Therefore, in order to avoid inelastic scattering from transitions between boundary states, we restrict our analysis throughout this work to weak coupling, and operate the pump in the limit $\beta_{\rm K}^{-1}, \Gamma^2/(\mu^2 \nu_0)  \ll\hbar/\beta, \hbar/T \ll \Gamma\ll\Delta$ where a scattering-matrix description is appropriate.  
[We remark that inelastic scattering may also occur due to transitions between (nearly degenerate) bulk ground states. However, the typical transition rate  for such processes are exponentially smaller than the edge excitations due to orthogonality of the many body ground states.  This is a manifestation of the fact that in 1+1 dimensions it is possible to spontaneously break a discrete symmetry~\cite{SColemanPRD1977}.  Hence, inelastic scattering arising from bulk degeneracies are absent in the above limit.]

We note that while a finite chemical potential is imperative to ensure the scattering matrix remains unitary, for the purpose of classification, we need only consider the particle hole symmetric part of the reflection matrix, $\tilde{\cal R}$~\cite{DMeidanPRB2010}. Hence, 
similar to their non interacting counterparts, interacting 
spin pumps may be classified by the topology of the  loop which 
$\tilde{\cal R}(t)$ forms on  $S^3$.

While the existence of a bulk energy gap ensures the description in terms of a reflection matrix remains valid, its many body nature gives rise to a richer variety of classes. 
Notably, interactions
can change the Fermi sea of non-interacting electrons 
into multiple many-particle 
ground states. The (near) degeneracy of these states is not protected by symmetry, and therefore split by small perturbations, such as the coupling to the leads, unlike in non abelian systems, which have a topologically non-trivial $ H(t)$ at all times. In the following we show that multiple ground states may alter both the periodicity and the time-reversal restriction given in Eq.~\eqref{niconstr}, thus modifying the classification of pumps.

To understand the implication of  $d$ nearly degenerate ground states, we note that the ground state of the macroscopic system spontaneously breaks the symmetry of the Hamiltonian.  The
reflection matrix will therefore depend on the specific ground state, $\phi_a$,  the system is prepared in at the beginning of the cycle, $\tilde{\cal R}_a(t)=\tilde{\cal R}[\phi_a(t)]$, typically the true ground state. The periodicity of the Hamiltonian ensures that the ensemble of ground states $\{\phi_a\}_{a=1,..,d}$  is restored after a period $T$, but the system need not return to the ground state $\phi_a$ it was in at $t=0$ (as discuss above, relaxation to the true ground state  occurs at times exponentially larger than $ T$). There are $d$  possible scenarios: After a full cycle of the pump, the system may either return to the original ground state $\phi_a$, or it may evolve to one of the $d-1$ other ground states. The latter scenario will result in an extended periodicity of the pump. In particular, the reflection matrix ${\cal R}_a(t)$  can have $d$ different periods $\tilde{\cal R}_a(t+nT)=\tilde{\cal R}_a(t)$,
where $ 1\leq n \leq d$. (Typically $n $ would divide $ d$, however, there is no fundamental reason why this should be the case). 

Due to the multiplicity of the ground state, the time-reversal restriction~\eqref{time-reversal transformations} applies to the ensemble of ground states only; it does not directly imply the relation~\eqref{niconstr} on the reflection matrix $\tilde{\cal R}_a$ of a particular physical realization. To find the corresponding time-reversal restriction of a general  $ nT$ periodic pump, we look at all the TRIM of the Hamiltonian during the extended cycle, $t_k=kT/2 $ for $0\leq  k\leq 2n$. If  the ground state is \textit{not} time-reversal symmetric at any of these points $\phi_a(t_k)\neq \Theta \phi_a(t_k)\Theta^{-1} $, the reflection matrix does not have any restrictions on the pumping cycle arising from time-reversal symmetry. Such a loop $\tilde{\cal R}_a(t)$  can be contracted onto a single point, and is therefore in the trivial class of pumps. Conversely, if there exists a point $t_k \equiv 0$ at which $\phi_a(0)=\Theta \phi_a(0)\Theta^{-1} $ then  $\phi_a(-t)=\Theta \phi_a(t) \Theta^{-1}$ and consequently $\tilde{\cal R}_a(-t) = \Theta\tilde{\cal R}_a(t)\Theta^{-1} $ for all $t $. 
Combined with the extended periodicity, $ nT$,  of the reflection matrix this ensures that the  loop  $\tilde{\cal R}_a(t)$  is restricted by exactly two TRIM, $t_i = 0,nT/2 $. 
The existence of  two TRIM allows one to distinguish two classes of loops for \textit{each} $nT$-periodic pump, Eq. \eqref{z2_index}:  A trivial pump which forms a  loop that can be contracted to a single point, and a non-trivial pump that completes an uncontractable loop after $n$ cycles of the pump. Hence, spin pumps with a $d$  ground states can  be classified by a  
$Z_{d+1}$ index
\begin{eqnarray}\label{zd+1_index}
z_{d+1}&=& n\, z_2\in \{0,1..,d\}
\end{eqnarray}
with $1\leq n\leq d $ and $z_2=0,1$. This index discerns a trivial class of pumps which do not pump quantized spin, from $d $ non-trivial classes.

\begin{figure}
\begin{center}
\includegraphics[width=0.45\textwidth]{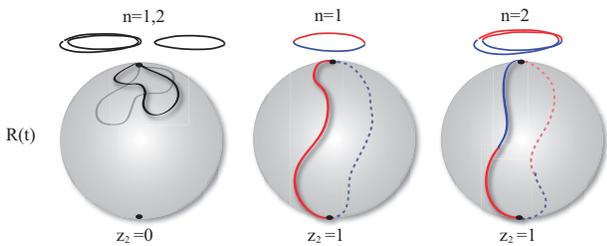}
\caption[0.5\textwidth]{The three different classes of pumps for systems with 2 ground states corresponding to (left to right): the trivial class with  $z_3 =n\times 0 = 0 $,  the integer pump with $z_3 = 1\times 1 $,  and the fractional pump with $z_3 =2\times 1 $. Here, the upper circles depict  the extended cycle in parameter space and for illustration purposes we restrict the reflection matrix to the two-sphere.   }
\label{class}
\end{center}
\end{figure}

{\it Pumping of quantized spin:---}
The $Z_{d+1} $ classification of the reflection matrix~\eqref{zd+1_index} has a direct effect on 
the spin pumped during a cycle.  In the weak coupling limit, the $d $ non-trivial pumps characterized by  $z_2=1 $ allow, in contrast to their trivial counterpart, for the noiseless pumping of quantized spin, 
even in the absence of a fixed spin quantization axis during the entire pumping 
cycle~\cite{DMeidanPRB2010}. The extended periodicity $ nT$, on the other hand, determines the averaged spin which is pumped during a cycle.

The class of non-trivial pumps which traverse a single resonance during the extended pumping cycle $nT $ can be described by a rotation around a fixed axis $\vec{e}_\varphi(t_i)$~\cite{DMeidanPRB2010}:
\begin{align}
\label{generic}
\tilde{\cal R}_a(t)=e^{i\varphi(t) \vec{e}_\varphi(t_i)\cdot \vec{\sigma}}.
\end{align}
Here $t_i=0$ or $nT/2$ is the TRIM at which the resonance occurs and $\vec{\sigma}$ are the spin Pauli matrices.
Consequently, the average spin per cycle $ T$ injected into lead $\alpha$ by these pumps is a fraction $1/n$ of the total spin pumped 
in the non-interacting case
\begin{align}
\label{pprop}
\langle \vec{S} \rangle_T 
=  {\hbar \over 2\pi n} \vec{e}_{\varphi}(t_i)
\int_0^{nT} dt \,\dot{\varphi}(t)=\frac{\hbar}{n} \vec{e}_{\varphi}(t_i). 
\end{align} 
Conversely, a trivial pump either remains insulating during the entire cycle, or traverses two resonances at the TRIM. In the weak coupling limit, the former group can be approximated by a constant reflection matrix $\tilde{\cal R}_a(t)\approx \openone $, and thus does not pump spin, while the latter cannot be described by a \textit{time independent} spin direction that would lead to a quantized spin pumped. 
The  $Z_{d+1}$ classification \eqref{zd+1_index} of interacting pumps  together with the average fractional spin pumped during a cycle \eqref{pprop} constitute the main result of this paper.

The extended periodicity of the pumping cycle in the fractional spin pumps  is reminiscent of the Aharonov-Bohm periodicity in  a ring made of a material in the fractional quantum Hall state at $\nu=1/3$: The ground state of the $\nu=1/3 $ is (nearly) threefold degenerate. Threading the ring by a single flux quantum $\phi_0 = h/e $ interchanges these ground states and  the system returns to its initial state after the flux changes by $ 3\phi_0$, giving rise to an Aharonov-Bohm periodicity of $ 3\phi_0$~\cite{RTaoPRB1984,QNiuPRB1985,DJThoulessPRB1989,XQWenPRB1990}.
We note that pumps constructed out of different physical systems with different ground-state degeneracies but with the same periodicity $nT$ pump the same average spin per cycle.  
This is related to the observation that  topological orders in fractional quantum Hall states cannot be characterized by the Hall conductance alone.

\begin{figure}[h!]
\begin{center}
\includegraphics[width=0.45\textwidth]{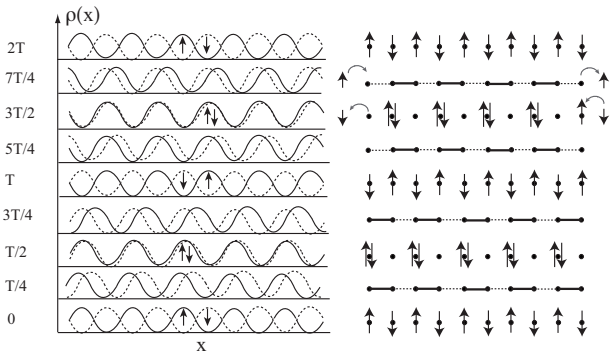}
\caption[0.5\textwidth]{An interaction driven pump described by the Hamiltonian in Eq. \eqref{example}  characterized by a $z_3 = 2$ index. The two  ground states at $t=0 $ and $t=T $ can be obtained from one another by creating a single spin flip in the bulk which then propagates to the edges of the wire.
The system returns to its original state after two pumping cycles. After a single pumping cycle the spin density waves for the the two orientations have shifted by half a wavelength, corresponding to the transfer of $1/2 [\hbar/2-(-\hbar/2)]$ spin from the left to the right edge of the system, without transferring charge.    }
\label{pump}
\end{center}
\end{figure}

{\it Example and bosonization:---}
As an example of  a spin pump characterized by $ z_3=2$, we study the  one-dimensional system with a half-filled energy band where  interactions give rise to a bulk gap,
\begin{eqnarray}\label{example}
H_{\rm int} &=&\sum_iU n_{i,\uparrow}n_{i,\downarrow}+\left(U/2-\delta V(t)\right) n_{i}n_{i+1} \\
&& \mbox{} + U_H(t)\left(n_{i,\downarrow}\psi_{i\uparrow}^\dag\psi_{i+1,\uparrow}-n_{i,\uparrow}\psi_{i\downarrow}^\dag\psi_{i+1,\downarrow}+\textrm{h.c.}\right). \nonumber
\end{eqnarray}
Here $ \delta V(t) =\delta V\cos{(2\pi t/T)}$, $U_H(t)=U_H\sin (2\pi t/T) $ set the strength of the time-dependent interaction terms in the Hamiltonian.
The ground state of this system adiabatically switches between spin and charge ordered insulating and interaction driven dimerized phases, see Fig~\ref{pump}. 
For $U_H =0 $, and $ \delta V(t)>0$ the electrons occupy different  sites and the system is in the spin density wave (SDW) ground state. At $ \delta V(t)<0$  the electrons pair up on the same site resulting in a charge density wave (CDW) ground state. The third interaction term $U_H $ breaks time-reversal symmetry and can e.g. be generated in the presence of a staggered magnetic field and alternating bond strength~\cite{CommentRelevantPert}. Similar models have been studied  in Ref.~\cite{SSarkarPRB2008,EBergCondMat2010,MMulliganCondMat2010}.

Interacting systems in one dimension are conveniently described in bosonization formalism. Here, the opening of the excitation gap arises due to the pinning of the bosonic degrees of freedom. In our example the pinning of spin degree of freedom $\phi_\sigma $ follows from the bosonized expression for $H_{\rm int}$,
\begin{eqnarray}
H_{int}[\phi_\sigma]&\sim&\delta V(t)\cos{4\phi_\sigma} +U_H(t)\sin4\phi_\sigma.
\end{eqnarray}
The four multiples of the bosonic phase  $4\phi_\sigma$ reflect the nature of the gap arising due to interaction terms containing four fermion operators. This results in a doubled periodicity compared to single particle gapped phases and consequently two ground states corresponding to the pinning of the bulk phase at $\phi_{\min}(t) $ and $\phi_{\min}(t)+\pi/2 $. 
In addition to the bulk pinning, the boundary condition imposed by coupling the wire by weak links to non interacting leads pins the bosonic phase 
at the edge of the wire. During the course of the cycle, the bulk pinning value is varied while the boundary pinning remains unchanged. As a result the bosonic phase develops a  kink close to the edge
giving rise to a resonance in the tunneling density of states at the
edge of the wire~\cite{OAStarykh2000}. 

Figure \ref{pump}  illustrates the ground state of the system described by Eq.\ ~\eqref{example}.  Due to the ground state multiplicity, the system returns to its original state after two pumping cycles. (A similar observation was made in Ref.~\cite{MMulliganCondMat2010}.) 
After a single pumping cycle the spin density waves for the the two orientations have shifted by half a wavelength, resulting in the transfer of a spin of $\hbar/2$ from the left to the right edge of the system, without transferring charge.

{\it Summary:---}We have studied the topological classification of one-dimensional insulators with a time-reversal restriction on the pumping cycle, in which the bulk excitation gap arises due to electron-electron interaction. 
Our classification holds in the weak coupling limit, where common sources of inelastic scattering can by avoided and a description in terms of a unitary scattering matrix is possible.
We found that a system with $d$ many body ground states can give rise to $d+1$ different classes of spin pumps. These include a trivial spin pump, which does not pump quantized spin, a quantized integer spin pump and $d-1$ fractional spin pumps, that allow for the average pumping of fractional spin $\hbar/n $ through the insulator. 
Recent works show that interactions may lead to new fractional topological insulators in the presence of a ground state degeneracy~\cite{JMaciejkoPRL2010,BSwingleCondMat2010}. The relation of our findings to the existence of a fractional quantum spin Hall state, remains an interesting question.

We gratefully acknowledge discussions with F.\ von Oppen and G.\ Zarand. This work is supported by the Alexander von Humboldt Foundation.


\begin{thebibliography}{10}
\bibitem{CommentAccResonance}
We disregard any accidental resonances that occur away from the TRIM, as they
  may be removed by a perturbation that preserves time-reversal symmetery.
\bibitem{CommentRelevantPert}
Such interaction terms are the relevant gap-opening perturbations in a
  Luttinger Liquid with interaction parameters $K_\sigma<K_\rho/3<1 $, as will
  be discussed in a future publication.
\end{thebibliography}
\end{document}